\documentclass[]{aa}
\usepackage{graphicx}
\begin{document}
\title{Possible flakes of molecular hydrogen in the early Universe}

\titlerunning{H$_2$ flakes in the early Universe}
\author{Daniel Pfenniger \inst{1} \and Denis Puy \inst{1,2}}

\institute{
  Observatory of Geneva, University of Geneva, CH-1290 Sauverny, Switzerland\\
  \email{Daniel.Pfenniger@obs.unige.ch} \and Institute of Theoretical
  Physics, University of Z\"urich,
  Winterthurerstrasse 190, 8057 Z\"urich, Switzerland\\
  \email{puy@physik.unizh.ch} } \offprints{Daniel Pfenniger,
  \email{Daniel.Pfenniger@obs.unige.ch}}

\date{Received 27 September 2002 / Accepted 3 November 2002}

\long\def\Htwo{H$_2$}

\abstract{The thermochemistry of \Htwo\ and HD in non-collapsed,
  non-reionized primordial gas up to the end of the dark age is
  investigated with recent radiation-matter and chemical reaction
  rates taking into account the efficient coolant HD, and the
  possibility of a gas-solid phase transition of \Htwo.  In the
  standard big-bang model we find that these molecules can freeze out
  and lead to the growth of flakes of solid molecular hydrogen at
  redshifts $z \approx 6-12$ in the unperturbed medium and under-dense
  regions.  While this freezing caused by the mere adiabatic cooling
  of the expanding matter is less likely to occur in collapsed regions
  due to their higher than radiation background temperature, on the
  other hand the super-adiabatic expansion in voids strongly favors
  it.  Later reionization (at $z \approx 5-6$) eventually destroys all
  these \Htwo\ flakes.  The possible occurrence of \Htwo\ flakes is
  important for the degree of coupling between matter and radiation,
  as well as for the existence of a gas-grain chemistry at the end of
  the dark age.
}

   \maketitle

\section{Introduction}
The period of time between the last scattering of the background
radiation photons and the formation of the first bound objects (i.e.,
the ``dark age'') still presents interesting unknowns even about the
main processes that happened then, as described below.

The studies of the physical and chemical conditions in the
post-recombination Universe clearly established that trace amounts of
molecules such as \Htwo, HD and LiH were present in the primordial gas
(Lepp \& Shull \cite{lep84}; Dalgarno \& Lepp \cite{dal87}; Latter \&
Black \cite{lat91}; Puy et al.~\cite{puy93}; Stancil et
al.~\cite{sta96}, \cite{sta98}; Galli \& Palla \cite{gal98}).  A
recent and comprehensive review of primordial chemistry is given by
Lepp et al.~(\cite{lep02}).

Because of its rotational levels, \Htwo\ is an important coolant of
interstellar gas from $10^4$ to $10^2\,$K.  Despite its lower
abundance, the molecule HD is also important, since its allowed dipole
rotational transitions provide strong channels coupling radiation and
matter at temperatures below $10^2\,$K, well below that of the lowest
\Htwo\ quadrupole transitions (at $512\,$K).  Thus primordial
molecules can play an important role in the post-recombination
Universe, already in the uniform, non-collapsed fraction of the
Universe.

Molecules can play an even more crucial role in the first collapsing
structures appearing at temperatures typically below a few hundred K,
since they offer the possibility to efficiently cool gaseous
proto-clouds (see Lahav \cite{lah86}; Puy \& Signore \cite{puy96},
\cite{puy97}; Uehara et al.~\cite{uhe96}, \cite{uhe00}; Abel et
al.~\cite{abe97}, \cite{abe00}).  However a definitive understanding
of the collapse of the first structures is still lacking, because the
coupling of gravity, chemistry, and radiation constitutes a formidable
non-linear system for which not all equations are presently well
known.  Since the system of equations is ``stiff'', i.e.,
exponentially sensitive to perturbations, changes in the conclusions
can eventually arise if a single factor is modified.  For instance
Combes \& Pfenniger (\cite{cp98}) show that the fraction of molecular
hydrogen can reach unity if the collapsing structures grow not as
smooth isolated spheres, but similarly to the fractal structures
observed in the galactic ISM.

Zwicky (\cite{zwi59}), Takeda et al.~(\cite{tak69}), Hirasawa et 
al.~(\cite{hir69}) and Matsuda et al.~(\cite{mat69}) pointed out early
that molecules could appear during the post-recombination period. The
mechanism by which molecules form in the post-neutralized Universe
is traditionally assumed to be different from the chemistry of the
interstellar medium, because the primordial chemistry is supposed to
remain a pure gas-phase chemistry.

The first natural extension of molecular physics toward larger matter
aggregates is the possibility of a gas-solid phase transition by
molecular hydrogen in the primordial gas%
\footnote{No corresponding $^4$He gas-solid transition exists.
  Below $5.2\,$K, the critical point of $^4$He, down to $0\,$K the
  $^4$He gas-liquid phase transition occurs at much higher pressure
  than for the \Htwo\ gas-solid transition.}.  
This is a central question, because the occurrence of \Htwo\ ice
grains, or more likely fluffy structures such as flakes, may radically
change the chemistry via grain surface reactions, as well as the
matter-radiation coupling via photon-grain interactions.

Solid molecular hydrogen%
\footnote{Not to be confused with metallic hydrogen, a high pressure
  form of hydrogen, found for example inside Jupiter.}, the simplest
molecular crystal, is rather special among the molecular crystals
because at low pressure the weakly anisotropic \Htwo\ molecules are
free to rotate and vibrate in the lattice.  Further, as for helium,
due to its large (comparable to the binding energy) zero-energy
motion, quantum effects are important; such crystals are known as
quantum crystals.  Solid hydrogen has been studied in the laboratory
for at least 70 years, and used in several cryogenic and industrial
applications, so many of its properties are today well established.  A
comprehensive review of solid molecular hydrogen is given by Silvera
(\cite{sil80}).

In this paper only first order effects associated with the condensed
phase of molecular hydrogen are considered.  Much more work will be
required to describe other possible effects occurring in the
non-linear regime following gravitational instability.  As it turns
out, much depends on the temperature of the matter, and whether matter
can be cooled in the interval $\sim 1-2\,$K by mere adiabatic
expansion.

Mechanical cooling is usually discarded because the overwhelming
number of photons in the cosmic background radiation is thought to
couple well enough matter and radiation with the few residual
electrons by Thomson scattering (see, e.g., the clear discussion by
Longair 1995. p. 428).  Here we calculate in detail the
thermo-chemical evolution with the most recent published rates when
the matter temperature cools as $\sim R^{-2}$, faster than the
radiation temperature $\sim R^{-1}$ due to the universal expansion
($R$ is the Universe scale factor).  The adiabatic cooling $\sim
R^{-2}$ is found to be effective below redshifts $\sim 300$, leaving
the possibility to cool non-collapsed, non-reionized matter to very
cold temperatures, until most of intergalactic matter is eventually
reionized.  Current models set the almost complete reionization around
redshifts $z \sim 5-6 $ (Razoumov et al.~2002).

The outline of this paper is as follows.  In Sect.~2, the method of
our calculations and the chemistry is described. In Sect.~3, results
about the revised thermochemistry are presented, including the
calculation of the phase transition to solid \Htwo.  The implications
are discussed briefly in Sect.~4.  Finally, our work is summarized in
Sect.~5.

\begin{table*}[h!btp]
 \renewcommand{\arraystretch}{1.0}
  \centering
    \begin{tabular}{|l|c||l|c|}\hline
    $H_0$ &  67 Km s$^{-1}$ Mpc$^{-1}$ & $Y_{\rm p}$ & 0.24\\ 
    $\Omega_{\rm r,0}$  &  $9.265 \times 10^{-5}$& [D] & $3.3 \times 
             10^{-5}$ \\ 
    $\Omega_{\rm m,0}$  &  $\sim 0.29990$ & [Li] & $2.1\times 10^{-10}$ \\ 
    $\Omega_{\rm b,0}$  &  0.0535   & [\Htwo ] & $10^{-40}$ \\ 
    $\Omega_{K,0}$  &   0 (flat Universe) & [HD] & $10^{-50}$ \\ 
    $\Omega_{\Lambda,0}$  & 0.7  & [LiH] & $10^{-50}$\\ \hline
    \end{tabular}
    \caption{\small{{\it Standard} cosmological model (S-model): 
parameters and initial relative abundances at $z=10^4$.}}
\end{table*}

\section{Model}
In most interstellar environments, association catalyzed on grain
surfaces is thought to dominate the formation of \Htwo. Thus,
molecular hydrogen is generally assumed to form by hydrogen atom
recombination on cold grains in dense molecular clouds (Hollenbach \&
Salpeter \cite{hol70}, \cite{hol71}). At the epoch of recombination
where a total absence of dust grains appears justified, surface
reactions are ignored.  The two standard and principal gas-phase
processes are characterized by 1) the H$_2^+$ process:
\begin{eqnarray}
  \rm H + H^+    & \longrightarrow&  {\rm H_2^+} + h\nu , \nonumber\\
  \rm H_2^+ + H  & \longrightarrow&  \rm H^+ + H_2 ,
\end{eqnarray} 
and 2) the H$^-$ process:
\begin{eqnarray}
  \rm H + e^-\   & \longrightarrow&  {\rm  H^-} + h\nu , \nonumber\\
  \rm H^- + H   & \longrightarrow&  \rm  e^-+ H_2 .
\end{eqnarray} 
However, these reactions are coupled with other reactions because the
main chemical species available from primordial nucleosynthesis are H,
H$^+$, D, D$^+$, He and Li. In particular this simple chemical network
must be coupled with the reactions which lead to the formation of HD
molecules, because HD is an efficient coolant below $\sim 100\,$K.

Recently Galli \& Palla (\cite{gal02}) gave an updated review of the
deuterium chemistry of the post-recombination Universe from their
exhaustive work on the primordial chemistry (Galli \& Palla
\cite{gal98}). The formation of HD in the primordial gas follows the
two routes:
\begin{eqnarray}
  \rm D^+ + H_2\   &\longrightarrow&  \rm H^+ +  HD , \nonumber\\
  \rm HD^+ + H     &\longrightarrow&  \rm H^+ + HD.
\end{eqnarray}

The radiative association
\begin{eqnarray}
  \rm H + D &\longrightarrow&  {\rm  HD} + h\nu .
\end{eqnarray} 
is very slow in our context.

The chemical reaction rates depend on the temperature of the species
or of radiation (for the photoprocesses), and of the density.  Thus
the chemical kinematics must be coupled with the hydrodynamic equation
in the framework of a cosmological model (here the Friedmann model).

A rigorous approach would be to consider \textit{four fluids}:
neutrals, positive ions, electrons and photons with their
corresponding temperatures in the hydrodynamic equations.  However
Flower \& Pineau des For\^ets (\cite{flo00}) showed a very small
difference between the temperature profiles of the neutrals, ions and
electrons in the primordial gas.  In fact, the coupling through
Coulomb scattering between the electrons and the ions and the neutrals
is sufficiently strong at the post-recombination epoch.  In the
following $T_{\rm m}$, is the common temperature of the neutrals,
ions, and electrons, which we call matter temperature.

The temperature of the cosmic background radiation (CBR) is given by:
\begin{equation}
  \frac{d T_{\rm r}}{d t} = - H(z) \, T_{\rm r},
\label{eq:tr}
\end{equation}
a relation which has been observationally confirmed by Srianand et
al.~(\cite{sri00}), at the redshift $z=2.138$, with absorption lines
from the first and second fine-structure levels of neutral carbon
atoms in an isolated cloud of gas.

$H(z)$ is the Hubble parameter at the redshift $z$, defined from the
scale factor $R$, and $H_0$ (the Hubble constant) is its present
value:
\begin{eqnarray}
 &&H(z)=   \dot{R} / R  = \\   
    && H_0 \, 
           \Bigl[
             \Omega_{\rm r,0} (1\!+\!z)^4 + \Omega_{\rm m,0} (1\!+\!z)^3 + 
                \Omega_{K,0} (1\!+\!z)^2 + \Omega_{\Lambda,0} 
           \Bigr]^{\frac{1}{2}},
\nonumber
\label{eq:h}
\end{eqnarray}
where the $\Omega_{,0}$ are constants with the present values,
constrained by the Friedman-Lema{\^\i}tre condition:
\begin{equation}
  \Omega_{\rm r,0} + \Omega_{\rm m,0} + \Omega_{K,0} + \Omega_{\Lambda,0} = 1.
\label{eq:fri}
\end{equation}
$\Omega_{\rm r,0}$, the radiation density parameter, is given by:
\begin{equation}
  \Omega_{\rm r,0} = \frac{a T_{\rm 3K}^4}{c^2} \, 
  \frac{8 \pi G}{3H_o^2} \, (1+f_\nu)
      \  \ {\rm with} \ \  f_\nu = \frac{21}{8} \times 
      \Bigr( \frac{4}{11} \Bigl)^{4/3},
\end{equation}
where $a=4\sigma/c$ is the radiation constant related to the
Stefan-Boltzmann constant $\sigma$, $T_{\rm 3K}=2.726\,$K is the
present radiation temperature (Mather et al.~\cite{mat94}). The
neutrino contribution to the radiation density for three massless
neutrino types is noted $f_\nu$, $G$ is the gravitational constant,
and $c$ is the speed of light. $\Omega_{\rm m,0}$ is the matter
density parameter (i.e., baryons and other non-baryonic matter),
$\Omega_{K,0}$ is the curvature parameter, and $\Omega_{\Lambda,0}$ is
the mass-energy density parameter associated with the Einstein's
cosmological constant $\Lambda$, or with quintessence (Wetterich
\cite{wet88}, Peebles \& Ratra \cite{pee02}).  In our context,
$\Omega_{\Lambda,0}$ will be held constant., fixed for $z$=0.

The evolution of the gas temperature $T_{\rm m}$ is governed by the
equation (see Puy et al.~\cite{puy93}):
\begin{equation}
  \frac{dT_{\rm m}}{dt} = -2 H(z) \, T_{\rm m} + 
             \Psi_{\rm Compton} + \Psi_{\rm mol} .
\label{eq:tm}
\end{equation}
$\Psi_{\rm Compton}$ characterizes the energy transfer from radiation
to matter via Compton scattering of CBR photons on electrons (Peebles
\cite{pee68}):
\begin{equation}
   \Psi_{\rm Compton} = 
      \frac{8}{3} \frac{\sigma_T a}{m_{\rm e} c} T_{\rm r}^4 
        \left(T_{\rm r} -T_{\rm m} \right) x_{\rm e} .
\end{equation}
$\sigma_T$ defines the Thomson cross-section and $m_{\rm e}$ the
electron mass.  $x_{\rm e}$ is the electron abundance (i.e., the
ionization fraction). $\Psi_{\rm mol}$ is the energy transfer via
excitation and de-excitation of molecular transition. However, in the
post-recombination context this thermal function is a heating function
and gives a slight contribution to the evolution of matter temperature
(see Puy et al.~\cite{puy93}).

This series of differential equations is completed by the redshift
equation:
\begin{equation}
  \frac{dz}{dt} = -H(z) \, (1+z) ,
\label{eq:z}
\end{equation}
and by the equation of the numerical density of baryons $n$:
\begin{equation}
  \frac{dn}{dt} = -3 H(z) \, n , 
\label{eq:n}
\end{equation}
which just describes the dilution of the conserved number of baryons
in the expanding volume.

Each chemical species $n_s$ has a similar behavior, completed with the
chemical network:
\begin{equation}
  \frac{dn_{\rm s}}{dt} = -3 H(z) \, n_{\rm s} + 
                    \Bigr( \frac{dn_{\rm s}}{dt} \Bigl)_{\rm chem},
\label{eq:ni_a}
\end{equation}
where the chemical network 
$\Bigr( \frac{dn_{\rm s}}{dt} \Bigl)_{\rm chem}$ 
has the form:
\begin{equation}
  \Bigr( \frac{dn_{\rm s}}{dt} \Bigl)_{\rm chem} \, = \,  
  \sum_{i,j} \, \alpha_{ij} n_i n_j -  \beta_{ij} n_{\rm s} .
\label{eq:ni_b}
\end{equation}
The first terms of the sum define the $s$ formation process ($i+j
\rightarrow s$) associated with the reaction rate $\alpha_{ij}$, the
second term characterizes the destruction process ($ s \rightarrow i +
j$) with the reaction rate $\beta_{ij}$.  Some reactions rates
$\alpha_{ij}$ and $\beta_{ij}$ are given in the Appendix.

\subsection{Initial conditions}
With stiff systems of equations, it is obviously important to use the
most accurate available physical constants and parameters. In the
context of the primordial chemistry the knowledge of the abundances of
the primordial nuclei is crucial.  From the latest determination of
primordial helium (i.e., $Y_{\rm p} \sim 0.24$) and deuterium ($\rm
D/H \sim 3.3 \times 10^{-5}$) abundances given by Tytler et
al.~(\cite{tyt00}) we can deduce the baryon density, which is in good
agreement with the cosmic microwave background measurements of
Boomerang and Maxima (de Bernardis et al.~\cite{ber00}, Jaffe et
al.~\cite{jaf00}): $\Omega_{\rm b,0} \, h^2 \, \sim \, 0.024$. These
experiments find a peak in the angular power spectrum of the microwave
background, consistent with that expected for cold dark matter models
in a flat Universe (i.e., $\Omega_{K,0}=0$).

The observations of type Ia supernovae (Perlmutter et
al.~\cite{per99}, Riess et al.~\cite{rie98}) suggest that the Universe
may be presently dominated by an additional {\it dark energy}.
Combined observations of type Ia supernovae, cosmic microwave
background anisotropy (Jaffe et al.~\cite{jaf00}) and cluster
evolution (Bahcall \& Fan \cite{bah98}) for which the results have
been done in the form of likelihood contours in the $\Omega_{\rm m,0}$
and $\Omega_{\Lambda,0}$ plane, yield
\begin{equation}
  \Omega_{\rm m,0}  \sim  0.3 \ \ {\rm and} \ \ \Omega_{\Lambda,0}  \sim  0.7.
\end{equation}
In our case $\Omega_{\rm m,0} \sim 0.2999$ and $\Omega_{\rm r,0} \sim
9.265 \times 10^{-5}$, in order to verify the Friedman-Lema{\^\i}tre
condition (Eq. \ref{eq:fri}).  We adopt the Hubble constant $H_0 = 67$
Km s$^{-1}$ Mpc$^{-1}$ given by Freedman (\cite{fre00}).  We summarize
the parameters of our \textit{standard} model (S-model) in Table 1.

In other models (see discussion) we will vary only the value of the
baryonic fraction $\Omega_{\rm b,0}$. Cosmochemistry evolution is not
sensitive to the $\Omega_{\Lambda,0}$ parameter. Molecular formation
appears in the redshift range $[100, 1000]$ whereas
$\Omega_{\Lambda,0}$ plays a \textit{role} in the Hubble parameter,
see Eq. \ref{eq:h}, at redshift below $z\sim 5$ (see Carroll, Press \&
Turner \cite{car92}).  We will fix this last parameter to the value
$\Omega_{\Lambda,0}=0.7$.

\begin{figure}
\vspace{-9mm}\includegraphics[scale=0.355,angle=-90]{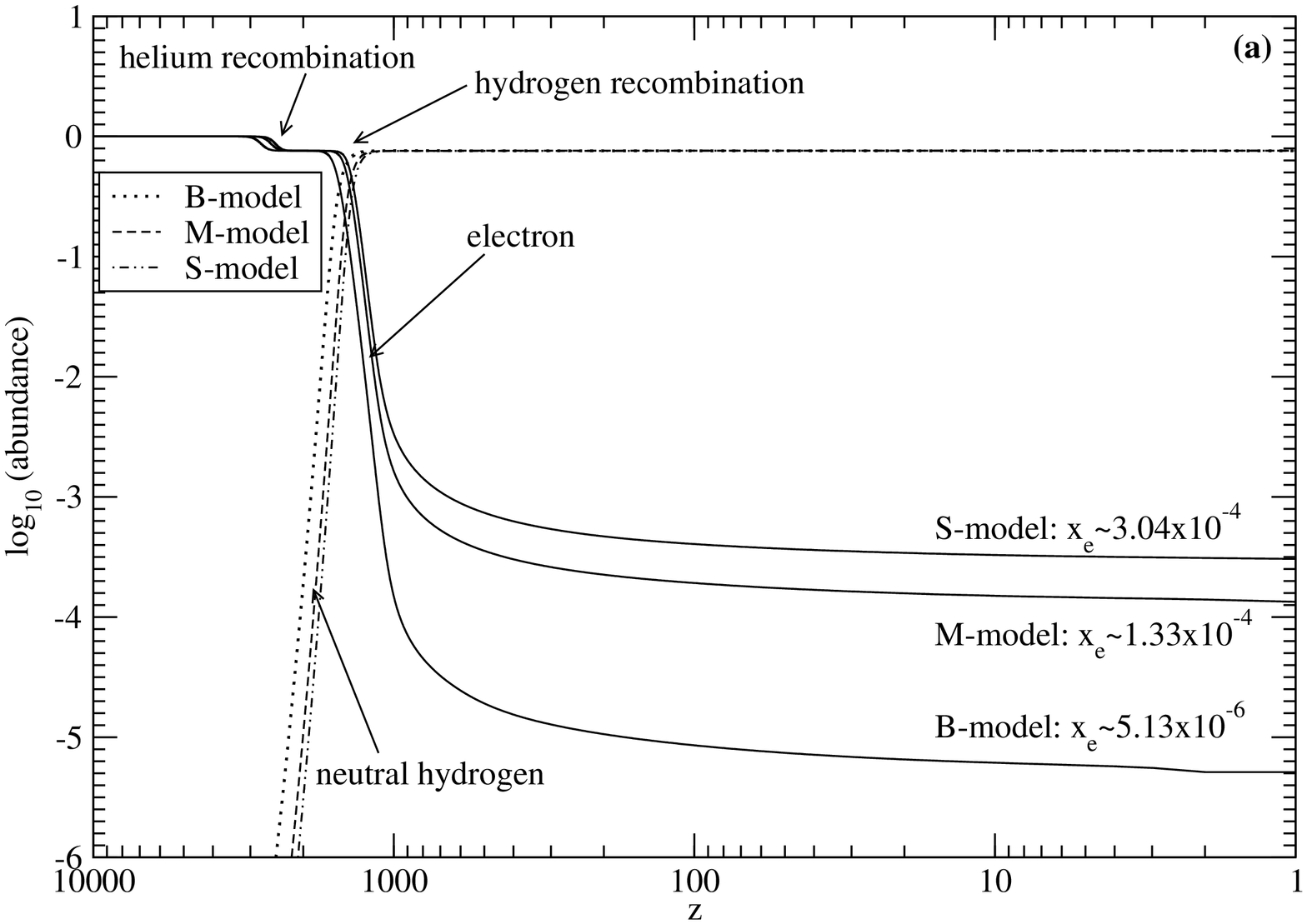}\vspace{-3mm}
\vspace{-3mm}\includegraphics[scale=0.355,angle=-90]{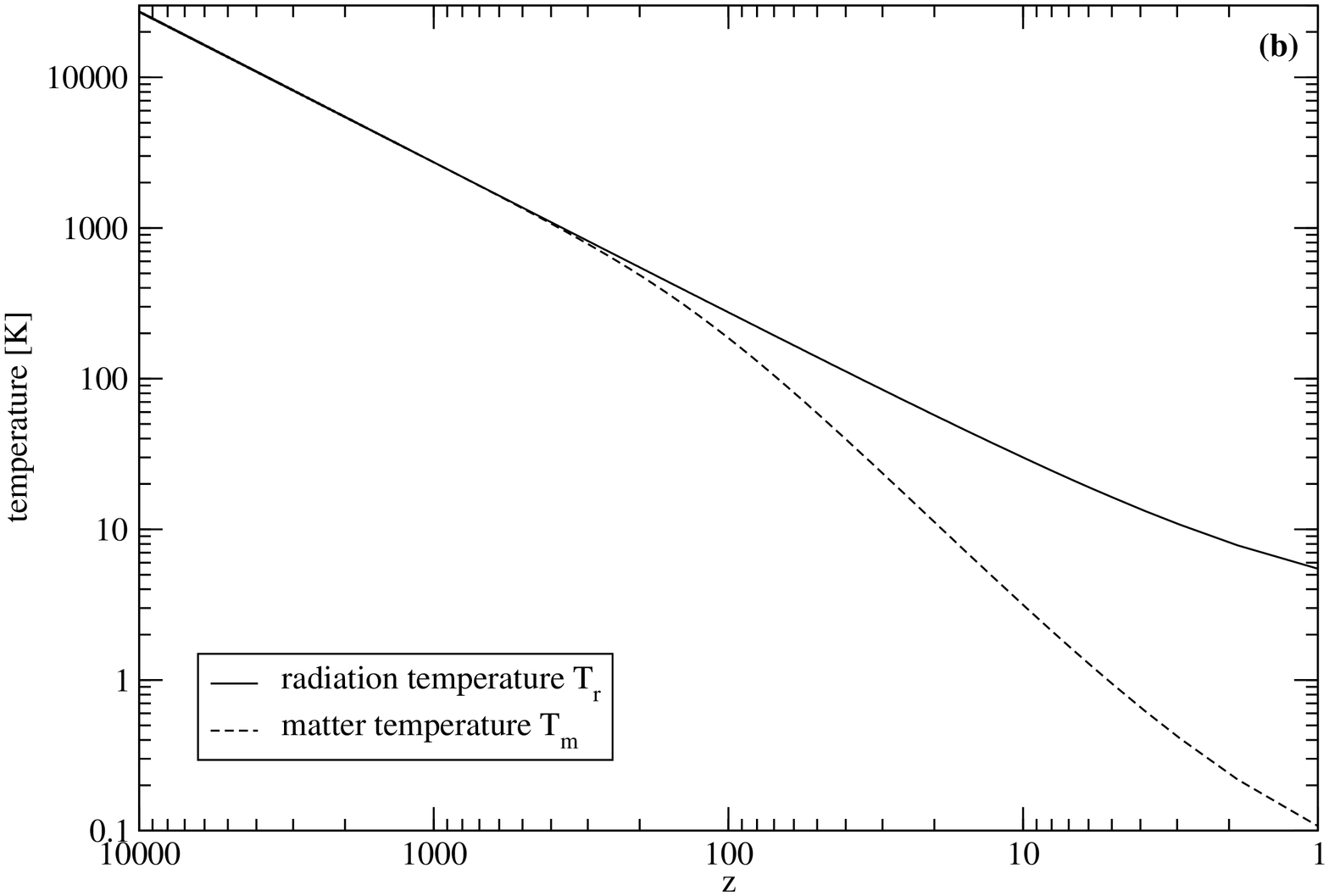}\vspace{-3mm}
\vspace{-3mm}\includegraphics[scale=0.355,angle=-90]{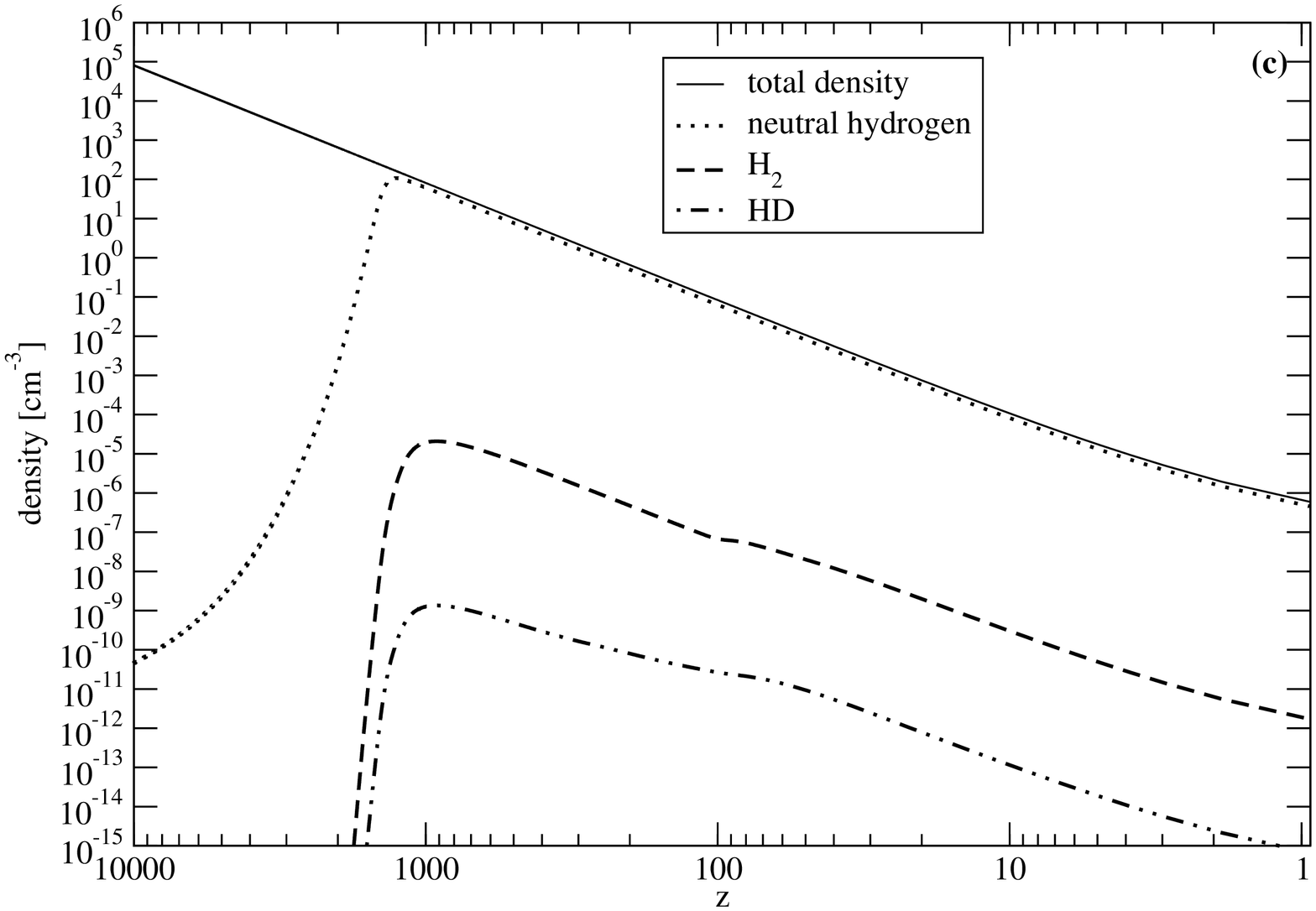}\vspace{-3mm}
\caption{ \textbf{a):} Helium and hydrogen recombination for a flat Universe 
  with $H_0=67\, \rm km\, s^{-1}\, Mpc^{-1}$ and
  $\Omega_{\Lambda,0}=0.7$.  The S-model corresponds to the baryonic
  fraction $\Omega_{b,0}=0.0535$, the M-model to $\Omega_{b,0}=0.1$,
  and the B-model to $\Omega_{b,0}=1$.  \textbf{b):} Thermal
  decoupling between radiation and matter in the S-model.
  \textbf{c):} Comparison of the \Htwo\ and HD density in the S-model.
  After a transient growth \Htwo\ and HD {\it follow} the evolution of
  the neutral density, which means that the fractional abundance
  becomes almost constant below $z \sim 500$.  }
\end{figure}

\subsection{Chemical evolution}
The numerical integration of the coupled chemical and hydrodynamical
equations is an initial value problem for stiff differential
equations. We use the chemical network developed by Galli \& Palla
(\cite{gal98}, \cite{gal02}), except the hydrogen and deuterium
recombination which is calculated from the reaction rates given by
Abel et al.~(\cite{abe97}), see Appendix.

We start our integration at redshift $z_i =10^4$.  At this value it is
particularly important to take into account%
\footnote{In our cosmological model, the equilibrium redshift (i.e.,
  between matter and radiation density) is $ 1+z_{\rm eq} =
  \Omega_{\rm m,0}/\Omega_{\rm r,0} = \frac{3H_{0}^2}{8 \pi G} \,
  \frac{c^2}{a T_{\rm 3K}^4}\, \frac{1}{1+f_\nu} \Omega_{\rm m,0} \sim
  3235$. This value justifies that it is necessary to consider the
  radiative component.}  the radiation contribution $\Omega_{\rm r,0}$
in order to calculate the age of the Universe $t_i$ at redshift $z_i$:
\begin{equation}
  t_i  = \frac{1}{H_0} \, \int_{1+z_i}^\infty \!
        \Bigr[ \Omega_{\rm r,0} x^4 + \Omega_{\rm m,0} x^3 
                           + \Omega_{K,0} x^2 + \Omega_{\Lambda,0} 
\Bigl]^{-1/2}
         \frac{dx}{x} 
\end{equation}
The time $t_i$ gives a timescale for the initial integration of the
set of Eqs.~(\ref{eq:tr} -- \ref{eq:ni_b}), which is solved by
Gear's method (Gear \cite{gea71}).

In Fig.~1(a) we show the helium and hydrogen recombination which give
the abundance of the free electrons. The slight decreasing of the
electronic fraction $x_{\rm e}$ at $z\sim 2485$ corresponds to the
helium recombination.  At the redshift $z\sim 1360$ the medium becomes
neutral (hydrogen recombination: neutral hydrogen is dominant). In all
of these cosmological models the fraction of electrons, $x_{\rm e}$,
is low at $z=1$.  For the S-model $x_{\rm e}=3.04 \times 10^{-4}$ is
in close accordance with the results of Galli \& Palla (\cite{gal98})
where $x_{\rm e} \sim 3.02 \times 10^{-4}$.  Before and around
recombination, the thermal history depends on the tight coupling
between radiation and matter resulting from Thomson scattering.  Well
after recombination, the free electrons \textit{quasi} vanished --i.e
$x_{\rm e} \sim$ few $[10^{-4}-10^{-6}]$, see Fig.~1(a).  The
evolution of the mean temperature of radiation and of matter is shown
in Fig.~1(b).  The subsequent evolution of matter temperature is
mainly caused by the dilatation of the Universe (first term in the
second member of Eq. \ref{eq:tm}), despite the very high numerical
photon density at low redshifts (see Longair \cite{lon95}, Partridge
\cite{par95}).  We notice that the thermal decoupling is effective at
redshift $z_{\rm dec} \sim 498$ (we define $z_{\rm dec}$ the redshift
below which the ratio $T_{\rm m}/T_{\rm r}$ is lower than 0.99). The
non-collapsed matter temperature can drop much below the radiation
temperature, unless some other coupling channels not included in our
model are effective. The possibility of cooling matter well below
radiation is crucial for the possibility to form \Htwo\ flakes.  In
Fig.~(1c) we have plotted the evolution of the density.  After the
thermal decoupling the total density is mainly due to neutral
hydrogen.  The \Htwo\ density rises through two processes: the first
step corresponds to the H$_2^+$ channel and the second one to the
H$^-$ channel. The HD density rises also in two steps with a
transition around $z \sim 100$, the (D$^+$, \Htwo) channel and the
(HD$^+$, H) channel.

\section{Thermochemistry}

Fifty years ago, van de Hulst (\cite{van49}) mentioned the possibility
of the existence of solid \Htwo\ in the interstellar medium, which
implies the possibility of a gas-solid phase transition.  This
possibility was elaborated by Wickramasinghe \& Reddish (\cite{wic68})
and Hoyle et al.~(\cite{hoy68}).  In the context of baryonic dark
matter as cold \Htwo\ Pfenniger \& Combes (\cite{pfe94}) and Wardle \&
Walker (\cite{war99}) found also that solid \Htwo\ can play an
important role.

This form of molecular hydrogen was already well known in the
laboratory for a long time.  For example, Harold Urey in his Nobel
prize lecture (1935) described how this phase transition was used to
easily separate the \Htwo\ and D$_2$ isotopes, thank to their slightly
different triple points.

From cryogenic industry tables (L'Air Liquide \cite{air76}), the
sublimation curve of para-hydrogen can be well fit with the following
equation (Pfenniger \& Combes \cite{pfe94}):
\begin{equation}
  p_{\rm sat} = 5.7 \times 10^{20} \, T_{\rm m}^{5/2} \, 
        {\rm exp}\biggl( -\frac{91.75}{T_{\rm m}} \biggr)  \ {\rm K\, cm^{-3}}
\label{eq:psat}
\end{equation}
in the range of matter temperature $T_{\rm m}$ between $1\,$K and the
triple point temperature at $T_3 = 13.8\,$K.  According to L'Air
Liquide (\cite{air76}) p. 890, below about $20\,$K over 99\% of normal
\Htwo\ is in the para form.  The asymptotic mixture containing the
maximum of ortho--\Htwo\ takes place between about 20 to $300\,$K,
above which the ortho/para ratio converges toward 3.  However, Flower
\& Pineau des For\^ets (\cite{flo00}), investigating the ortho/para
\Htwo\ ratio, found that this ratio decreases to a frozen value of
about $0.25$ for $z<20$ due to the low density of the main conversion
agent, H$^+$.  We will see such a different ortho/para ratio has a
very small effect in the result about the \Htwo\ condensation
redshift.

\begin{figure}
\vspace{-9mm}\includegraphics[scale=0.355,angle=-90]{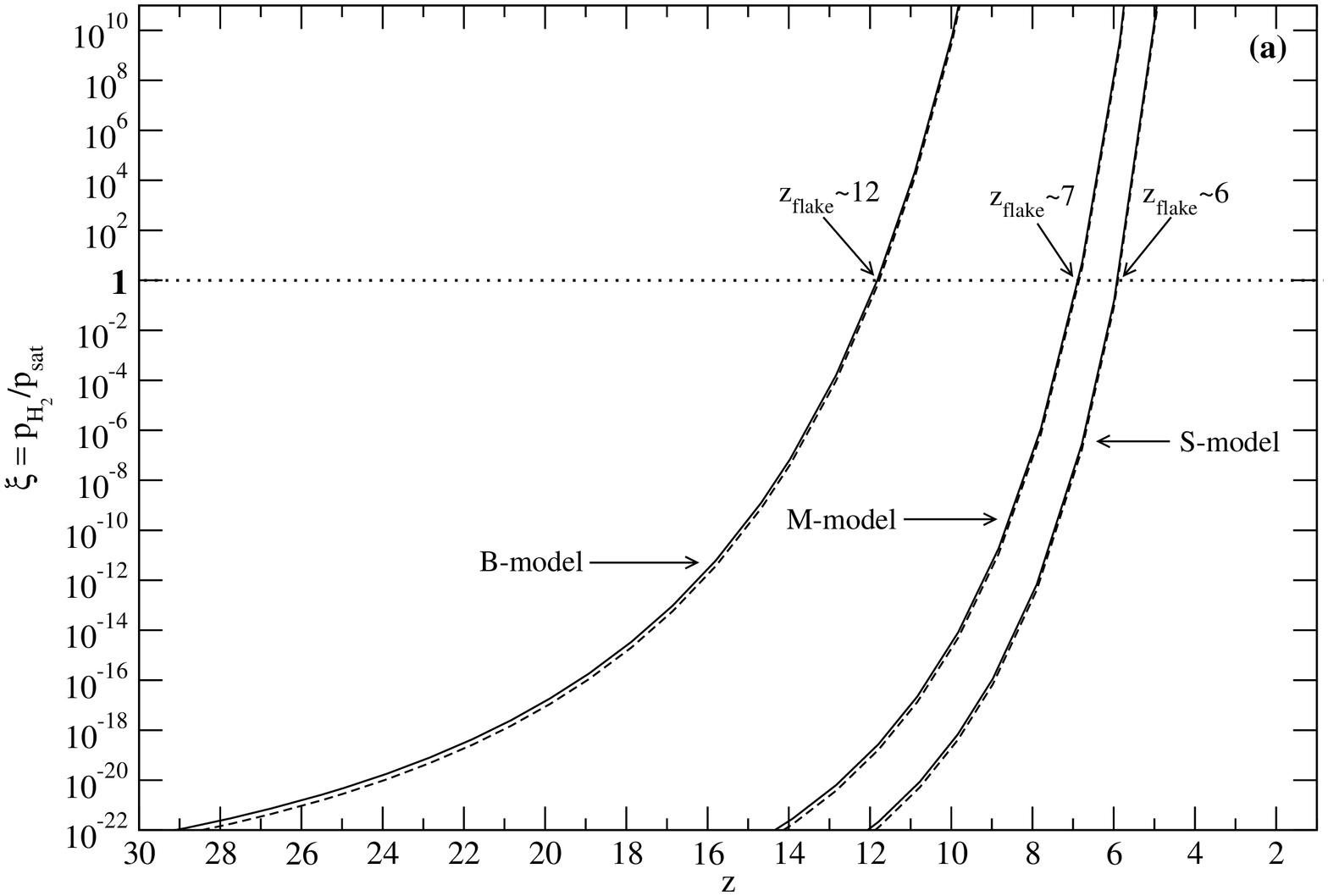}\vspace{-3mm}
\vspace{-3mm}\includegraphics[scale=0.355,angle=-90]{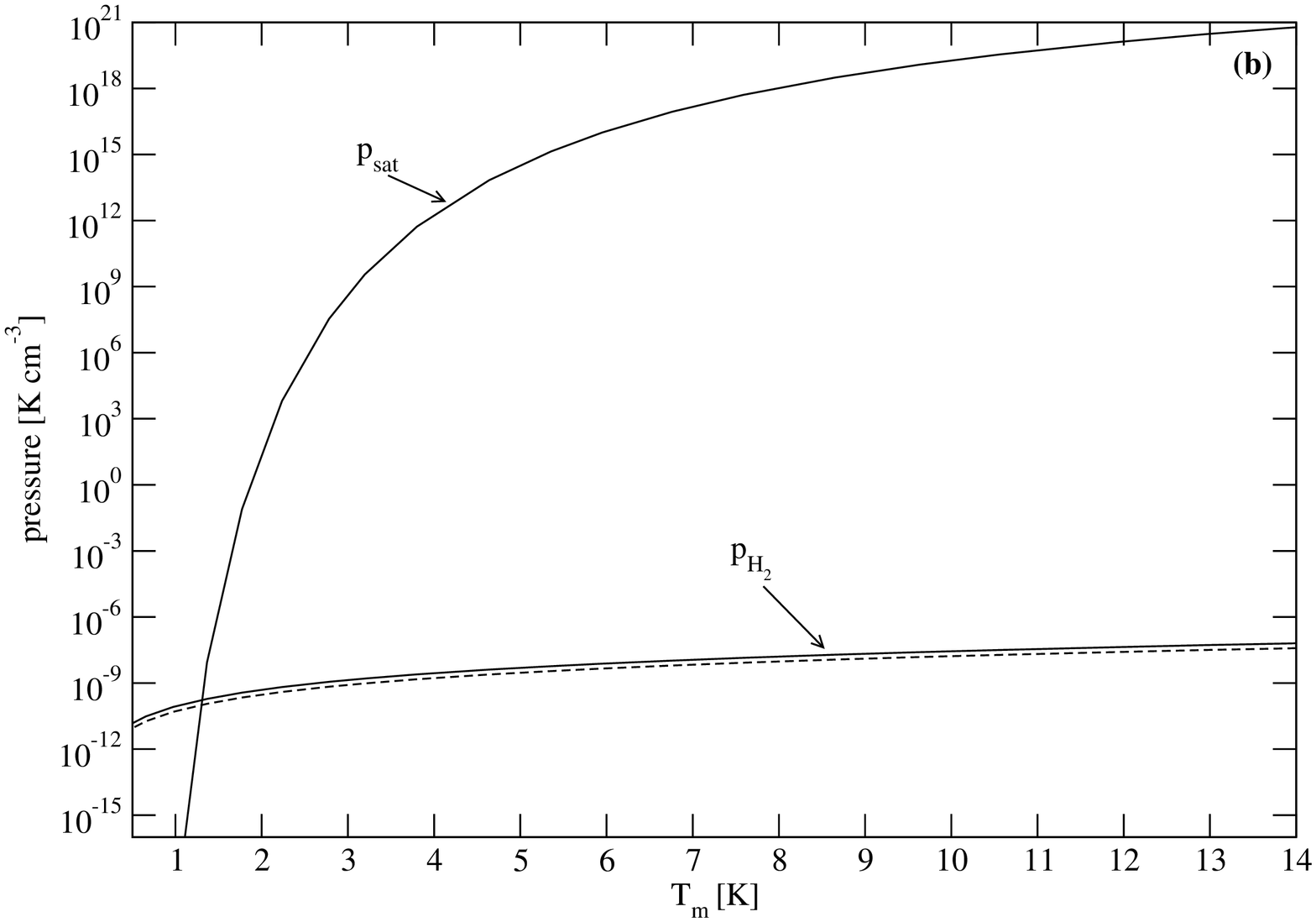}\vspace{-3mm}
\caption{
  \textbf{(a):} Comparison of the ratio between the gas pressure and the
  saturation pressure $p_{\rm sat}$, for \Htwo-HD gas between $1\,$K
  and $T_3 = 13.8\,$K (i.e., range of redshift between 1 and 30),
  $z_{\rm flake}$ is the redshift below which flakes form; 
  B-model: $\Omega_{\rm b,0} \sim 1$, 
  M-model: $\Omega_{\rm b,0} \sim 0.1$, 
  S-model: $\Omega_{\rm b,0} \sim 0.0535$
  (for all of these models $\Omega_{\Lambda,0}=0.7$). 
  For each model, the solid curves corresponds to molecules with 
  allowed rotation, while the dashed curves to non-rotating molecules.
  \textbf{b):} Comparison of the saturation pressure $p_{\rm sat}$ with the
  \Htwo\ partial pressure $p_{\rm H_2}$, for the S-model as a function 
  of temperature.
}
\end{figure}

If we suppose an ideal (\Htwo\ --HD) gas
\footnote{In this Sect., {\rm $p_{\rm H_2}= \frac{3}{2} n_{\rm H_2}
    T_{\rm m}$} where we neglect the rotation and vibration of the
  molecule, a valid approximation in the \Htwo\ solidification range
  $0-13.8\,\rm K$.},
the ratio $\xi$ between the \Htwo-gas pressure and the saturation
pressure becomes:
\begin{equation}
  \xi=\frac{p_{\rm H_2}}{p_{\rm sat}} \approx 2.63 \times 10^{-21} \, n_{\rm H_2} 
          T_{\rm m}^{-3/2} \exp\biggl( \frac{91.75}{T_{\rm m}} \biggr),
\end{equation}
where $n_{\rm H_2}$ is the \Htwo\ density (in cm$^{-3}$, see
Fig.~1(c)).  Although HD has a dipole moment when \Htwo\ has
quadrupolar transitions, the thermodynamics of HD gas is assumed to be
very close to \Htwo\ (without distinction between \Htwo\ and HD).

In Fig.~2(a) the ratio $\xi$ is plotted for three different
cosmological models: 1) S-model: the \textit{standard} cosmological
model, 2) M-model: model with $\Omega_{\rm b,0}=0.1$, 3) B-model:
model with $\Omega_{\rm b,0}=\Omega_{\rm m,0} \sim 0.3$.  For all of
these models $\Omega_{\Lambda,0}=0.7$).  We define $z_{\rm flake}$ as
the redshift for which below this boundary the ratio $\xi$ is greater
than 1, and thus leads to the possibility to freeze the (\Htwo--HD)
gas.

We find that $z_{\rm flake} \sim 12$ for the B-model, $z_{\rm flake}
\sim 7$ for the M-model, and $z_{\rm flake} \sim 6$ for the S-model.
The redshift $z_{\rm flake}$ is larger in the B-model because the
intrinsic baryon density is more important.

In Fig.~2 we show also the very slight influence of the molecule
rotation factor, $5/2$ or $3/2$, in the \Htwo~pressure.  The redshift
$z_{\rm flake}$ is not essentially affected.  The same remark is valid
when the ortho/para ratio is taken as 0.25 instead of 0, since $p_{\rm
  H_2}$ is lowered by only a factor 0.75.  In Fig.~2(b) we compare the
pressure $p_{\rm sat}$ and the partial pressure $p_{\rm H_2}$
(S-model) below the triple point.  Obviously, the insensitivity of
$z_{\rm flake}$ to variations of $p_{\rm H_2}$ results from the
steepness of the \Htwo\ sublimation curve.

\begin{figure}
\vspace{-9mm}\includegraphics[scale=0.355,angle=-90]{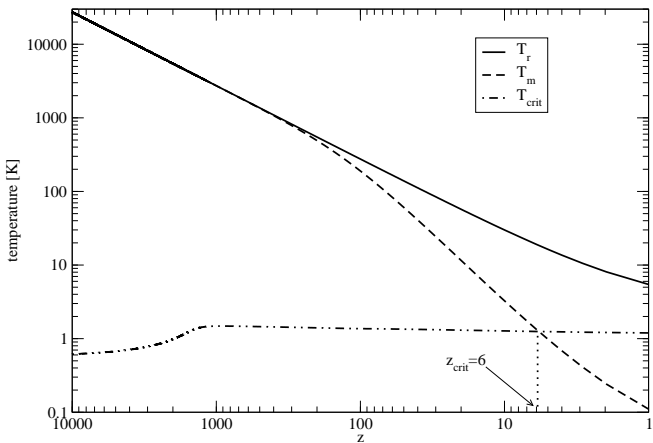}\vspace{-3mm}
\caption{Evolution of temperatures of matter and radiation. 
  The critical temperature is reached at redshift 6; below this
  redshift \Htwo\ could stick to the surface of ice.}
\end{figure}

\section{Discussion}

What we have shown above is that below a critical redshift a necessary
condition to form \Htwo\ solid grains is fulfilled.  However, this
condition is not sufficient, in reality several complications may
prevent or favor the formation of solid \Htwo.  We discuss below some
of these complications, as well as some consequences in case solid
\Htwo\ does form.

\subsection{Coupling of chemistry and \Htwo\ flakes}
Sandford \& Allamandola (\cite{san93}) showed experimentally that
\Htwo\ molecules can stick to the surfaces of usual ices, and argued
that gas-phase \Htwo\ freezes out onto classical dust grains. Thus
\Htwo-containing ices may be fairly common in dense molecular clouds.

On the other hand, solid hydrogen is a quantum solid with a large
zero-point energy (Silvera \cite{sil80}), which means that contrary to
usual dust grains, atoms or molecules adsorbed on the surface of
\Htwo\ ice do not stick at fixed lattice sites but move easily around
to go to the lowest free energy configuration.  The consequence is
that surface chemical reactions may be thus greatly accelerated on
\Htwo\ ice.

Atoms or molecules stick on the \Htwo\ ice surface below a critical
temperature which is estimated by Sandford \& Allamandola:
\begin{equation}
  T_{\rm crit} \, = \, \frac{100}{56.46-\ln(n_{\rm H_2})}\, {\rm K} .
\label{eq:crit}
\end{equation}

In Fig.~3, we compare the evolution of the radiation and matter
temperatures with the evolution of the critical temperature.  We see
that the critical redshift $z_{\rm crit} = 6$, i.e., also the critical
redshift for which the temperature of (\Htwo--HD) gas is below the
critical temperature (Eq. \ref{eq:crit}).  The value of $z_{\rm crit}$
is remarkably close to the threshold redshift leading to the formation
of \Htwo\ flakes (Figs.~2(a) and 2(b)), which is not unexpected since
condensation and sticking phenomena are both closely related to the
inter-molecular force.

Thus, we can imagine that once the first flakes form, the formation of
\Htwo\ and HD may be enhanced by neutral hydrogen reactions on the
flake, which in turn increase the \Htwo\ vapor pressure, increasing
the flake growth.

Sandford \& Allamandola (\cite{san93}) speculated on the possibility
that pure \Htwo\ ices might exist in intergalactic and extragalactic
space. This last hypothesis had received some attention previously
(Lee et al.~\cite{lee71} or Hegyi \& Olive \cite{heg86}).

\subsection{Coupling of background radiation and \Htwo\ flakes}
However, in the still neutral and not collapsed regions of the
Universe, the background radiation is warmer than the adiabatically
expanding matter.  If \Htwo\ flakes form, a new channel opens for
coupling matter and radiation, since solids possess numerous phonon
bands able to interact with infrared radiation.

The most prominent absorption bands in solid \Htwo\ occurs in the
interval $1.89-2.50 \, \rm \mu m$ (e.g., Allin et al.~\cite{all55}),
and at higher frequency in the interval $1.06-1.24 \, \rm \mu m$
(Varghese et al.~\cite{var87}).  They are in part caused by the
vibrational and rotational quadrupolar absorption as for the isolated
molecule, and by the main vibrational mode between two adjacent
molecules in the lattice. 

A particularity of solid \Htwo\ is that the molecules in the crystal
at low pressure are almost free to vibrate and rotate, molecular
transitions of low quantum numbers being allowed.  Therefore it may be
useful to mention a few other transitions for free molecules in the
far infrared.  Zwicky (1959) mentioned a line at $85 \, \rm \mu m$
associated to the \Htwo\ ortho-para transition, but this transition is
very slow, of order of 300 yr.  Particularly interesting for the dark
age context is the HD dipolar rotational transition around $114 \, \rm
\mu m$ (e.g., Lee et al.~\cite{lee88}) which, despite a low HD
abundance, is relevant since the transition is dipolar, and the
background radiation particularly intense in the far infrared.

Thus any coupling of radiation with solid \Htwo\ should increase the
matter temperature, which acts to sublimate \Htwo\ flakes, a
self-destructive process.  At any redshift an equilibrium temperature
should be reached where the condensation and sublimation rates are
equal.  Several uncertainties prevent us from estimating these
time-dependent rates, such as the radiation-flake coupling which
depends on the size, geometry and crystalline structure of the \Htwo\ 
flakes, and also the chemistry coupling mentioned above.  More
``cosmological'' laboratory experiments simulating aspects of the
physico-chemical conditions during the dark age would be useful.

\subsection{Heating from latent heat} 
The formation of \Htwo\ ice liberates molecular binding energy in the
form of heat, the latent heat.  In the range $1-3\,$K the latent heat
amounts to $92-99\,\rm kcal\, kg^{-1}$ (L'Air Liquide 1976), or
$22-62$ times the molecule kinetic energy $\frac{3}{2} kT$.  Since the
\Htwo\ abundance is then of the order of $10^{-6}$, clearly the latent
heat can easily be shared with the neutral gas without causing any
significant temperature increase.
  
\subsection{Destruction of \Htwo\ flakes by reionization}
At redshifts $5-9$, one expects that a growing fraction of the
non-collapsed Universe is filled by the UV radiation produced by the
first collapsed and light emitting structures, stars and quasars (see,
e.g., the models of Razoumov et al.~\cite{raz02}).  This means that at
redshift $6-7$ possibly most of the diffuse gas has already passed
through a ionization front, and is reheated to high temperatures.  So
the formation of flakes may be partly or totally suppressed if the
reionization occurs early enough, and propagates in every corner of
the Universe.  Voids are the last regions to be reionized, for reasons
well discussed by Razoumov et al.~(\cite{raz02}): in brief, UV
radiation percolates through ionization fronts and thus ionizes the
dense regions first.  The number of UV photons necessary to reach and
ionize voids is therefore much larger than, as earlier assumed, for a
UV optically thin medium bathed by a general radiation field.  The
result is to delay the full reionization of the Universe, leaving some
more time for \Htwo\ freezing, particularly in voids.
 
\subsection{Formation of \Htwo\ flakes in voids}
An additional mechanism enhancing \Htwo\ freezing occurs in regions
expanding faster than the Universe, i.e., in voids, since there the
adiabatic cooling is even stronger.  The essential factor favoring
\Htwo\ freezing is the temperature.  If the void matter reaches a
temperature around $1-2\,$K the formation of \Htwo\ flakes is very
much facilitated because the saturation pressure drops extremely fast,
by over 27 dex in this range, as can be derived from Eq.
~\ref{eq:psat}, or Fig.~2(b).  Since in an adiabatic mono-atomic gas $T
\propto \rho^{\gamma-1}$, a region that is 10 times underdense with
respect to the average would be, for $\gamma=5/3$, $10^{2/3}=4.6$
times colder.  Since $T_{\rm m} \propto (1+z)^2$, the redshift at
which \Htwo\ reaches the critical freezing temperature would about
double, bringing it at a much darker epoch $z>12$.
 
\subsection{Formation of \Htwo\ flakes in collapsing structures}
On the other hand, if gas radiative cooling $\propto n^2$ is
sufficiently strong, in collapsing regions eventually the temperature
increase by adiabatic compression is canceled by efficient radiative
cooling.  Primordial molecules such as HD are particularly efficient
for cooling at temperatures below $100\,$K.

The prediction of the temperature runs in gravitational collapsing
structures is highly non-trivial since the negative specific heat of
self-gravitating systems may lead to opposite behaviours than those in
familiar thermal systems.  Observations of the galactic ISM show that
unstable gas tends to adopt a fractal structure over $4-5$ orders of
magnitude, and the smallest gas clouds may remain very cold and dense.
In this context, Pfenniger \& Combes (\cite{pfe94}) argue that cold
\Htwo\ gas may be a candidate for baryonic dark matter provided that
the smallest elements of the gas fractal structure, the clumpuscules,
are sufficiently cold and compact.  In their postulated conditions set
essentially by the virial theorem, at low redshifts ($T \approx 3
\,$K, $n \approx 10^9 \,\rm cm^{-3}$) they noted that the conditions
are met to form solid \Htwo\ due to a much higher gas pressure than
the one contemplated above in the unperturbed Hubble flow.  However if
the cold collapsed structures reach the CBR temperature from above,
clearly condensed \Htwo\ (in solid or liquid forms) cannot form below
the redshift at which the radiation is at the \Htwo\ critical point
temperature, at $33.4\,$K.  In other words, the largest redshift for
the occurrence of condensed \Htwo\ in equilibrium with the CBR in
dense but cold clumps is at $z=33.4/2.726 -1 = 11.25$.

\section{Conclusions}
In the current understanding of the formation of the first bound
structures during the dark age, the possibility that solid hydrogen
flakes exist and modify the subsequent evolution must be considered.
For regions in which the growth of inhomogeneities can be neglected,
in widely different Universe models our chemical calculations show
that between redshifts 6 and 12 \Htwo\ flakes may start to form.  The
subsequent chemistry and matter-radiation coupling should be
significantly altered.  Since several factors that can either damp or
increase the formation of \Htwo\ flakes, clearly one must take into
account that the late dark age thermochemistry is likely to be much
more complex that earlier admitted.

However at these redshifts most of the first structure formation
scenarios predict an already well advanced and widespread stage of
reionization, which casts uncertainties about which processes win, or
which fraction of the Universe between voids and collapsed regions are
concerned by \Htwo\ flake formation.  The appearance of snowflakes at
the end of the dark age would constitute the first ``cosmic winter''.

\begin{acknowledgements}
  Part of this work was supported by the Tomalla-Stiftung and the
  Swiss National Science Foundation. The authors gratefully
  acknowledge useful discussions with F. Combes, M. Signore, and F.
  Melchiorri.  Part of the preparation of this manuscript has been
  done during a stay at the Aspen Center for Physics.
  We are grateful to the referee, Daniele Galli, for constructive comments. 
\end{acknowledgements}

\section*{Appendix}
\appendix
We use the reactions rates $\alpha_{ij}$ and $\beta_{ij}$ and
photo-processes given by Galli \& Palla (\cite{gal98}, \cite{gal02}),
except the hydrogen and deuterium recombination:
\begin{eqnarray}
   \rm H^+ + e^-  &\rightarrow& \rm H + \gamma  \nonumber \\
   \rm D^+ + e^-  &\rightarrow& \rm D + \gamma, 
\end{eqnarray}
with a reaction rate $\alpha_{\rm rec}$ (in cm$^{-3}$ s$^{-1}$), which is 
calculated from the approximation given by Abel et al.~(\cite{abe97}):
\begin{eqnarray}
  & \alpha_{\rm rec} \, = \, {\rm exp}\Bigr[& -28.6130338 - 0.72411256 \, 
   \ln(T_{\rm m}) \nonumber  \\
  &&-2.02604473 \times 10^{-2} \, \ln^2(T_{\rm m}) \nonumber \\
  &&-2.38086188 \times 10^{-3} \, \ln^3(T_{\rm m}) \nonumber \\
  &&-3.21260521 \times 10^{-4} \, \ln^4(T_{\rm m}) \nonumber \\
  &&-1.42150291 \times 10^{-5} \, \ln^5(T_{\rm m}) \nonumber \\
  &&+4.98910892 \times 10^{-6} \, \ln^6(T_{\rm m}) \nonumber \\
  &&+5.75561414 \times 10^{-7} \, \ln^7(T_{\rm m}) \nonumber \\
  &&-1.85676704 \times 10^{-8} \, \ln^8(T_{\rm m}) \nonumber \\
  &&-3.07113524 \times 10^{-9} \, \ln^9(T_{\rm m}) \Bigl] ,
\end{eqnarray} 
where $T_{\rm m}$ is in eV. We consider also the reverse reaction (i.e.,
photoionization of neutral hydrogen and neutral deuterium) :
\begin{eqnarray}
  \rm H + \gamma &\rightarrow&  \rm H^+ + e^- \nonumber \\
  \rm D + \gamma &\rightarrow&  \rm D^+ + e^- 
\end{eqnarray}
with the radiative rate coefficient $\beta_{\rm ph}$ (in s$^{-1}$) through 
the CBR, defined by:
\begin{equation}
  \beta_{\rm ph} \, = \, \frac{8 \pi}{c^2} \, 
  \int_{\nu_{\rm th}}^\infty \, \sigma_{\rm ph}(\nu) \, 
  \frac{\nu^2 \, {\rm d}\nu}{\exp( h\nu  / k T_{\rm r} ) - 1},
\end{equation}
where $\nu_{\rm th}$ is the threshold energy for which photoionization
is possible (here $h\nu_{\rm th}=13.6$ eV).  We take the frequency
dependent cross section $\sigma_{\rm ph}$ (in cm$^2$) given by Abel et
al.~(\cite{abe97}):
\begin{eqnarray}
& \sigma_{\rm th} = 
 & 6.30 \times 10^{-18} \Bigr( \frac{\nu}{\nu_{\rm th}} \Bigl)^4 \  
     \frac{{\rm exp} \Bigl[4-4 \arctan(\epsilon)/\epsilon \Bigr] }
       {1-\exp (-2\pi/\epsilon)} \nonumber \\
 & &   {\rm where} \ \epsilon =\sqrt{ \frac{\nu}{\nu_{\rm th}} -1} 
\end{eqnarray}



\begin{thebibliography}{}
  
\bibitem[1997]{abe97} Abel, T., Anninos, P., Zhang, Y., \& Norman, M.
  1997, New Ast. 2, 181

\bibitem[2000]{abe00} Abel, T., Bryan, G., \& Norman, M. 2000, ApJ
  540, 39
  
\bibitem[1955]{all55} Allin, J.E, Hare, W.F.J., \& MacDonald, R.E.
  1955, Phys. Rev 98, 554

\bibitem[1998]{bah98} Bahcall, N., Fan, X. 1998, ApJ 504, 1
  
\bibitem[1999]{bur99} Burles, S., Nollett, K., \& Turner, M. 1999,
  astro-ph/9903300

\bibitem[1992]{car92} Carroll, S., Press, W., \& Turner, E. 1992,
  Annu. Rev. Astr. Astrophys. 30, 499

\bibitem[1998]{cp98} Combes, F., \& Pfenniger, D. 1998, Mem. della
  Soc. Astron. Italiana, 69, 413

\bibitem[1987]{dal87} Dalgarno, A., \& Lepp, S. 1987, in: Molecular
  Astrophysics, Hartquist T.W. ed., Cambridge University Press, p473

\bibitem[2000]{ber00} De Bernardis, P., et al.~2000, Nature 404, 955
  
\bibitem[2000]{flo00} Flower, D., \& Pineau des For\^ets, G. 2000,
  MNRAS 316, 901

\bibitem[2000]{fre00} Freedman, W. 2000, Phys. Report 333, 13

\bibitem[2000]{jaf00} Jaffe A., et al.~2000, astro-ph/0007333

\bibitem[1998]{gal98} Galli, D., \& Palla, F. 1998, AA 335, 403

\bibitem[2002]{gal02} Galli, D., \& Palla, F. 2002, astro-ph/0202329
  
\bibitem[1971]{gea71} Gear, W. 1971, in: Numerical initial value
  problems in ordinary differential equations, Prentice-Hall eds

\bibitem[1986]{heg86} Hegyi, \& D., Olive, K. 1986, ApJ 303, 56
  
\bibitem[1969]{hir69} Hirasawa, T., Aizu, K., \& Taketani, M. 1969,
  Prog. Theor. Phys. 41, 835

\bibitem[1970]{hol70} Hollenbach, D., \& Salpeter, E. 1970, J. Chem.
  Phys. 53, 79

\bibitem[1971]{hol71} Hollenbach, D., \& Salpeter, E. 1971, ApJ 163,
  155
  
\bibitem[1968]{hoy68} Hoyle, F., Wickramasinghe, N., \& Reddish, V.
  1968, Nature 218, 1124

\bibitem[1986]{lah86} Lahav, O. 1986, MNRAS 220, 259
  
\bibitem[1976]{air76} L'Air Liquide, 1976, Gas Encyclopedia, Elsevier
  Eds
  
\bibitem[1991]{lat91} Latter, W., \& Black, J. 1991, ApJ 371, 161
  
\bibitem[1988]{lee88} Lee, S.Y, Lee, S., \& Gaines, J.R. 1988, Phys.
  Rev. B, 37, 2357

\bibitem[1971]{lee71} Lee, T., Gowland, L., \& Reddish, V. 1971,
  Nature Phy. Sci. 231, 193

\bibitem[1984]{lep84} Lepp, S., \& Shull, M. 1984, ApJ 280, 465
  
\bibitem[2002]{lep02} Lepp, S., Stancil, P., \& Dalgarno, A.
    2002, J. Phys. B 35, R57
  
\bibitem[1995]{lon95} Longair, M. 1995, in: The Deep Universe,
  Saas-Fee Advanced Course 23, Eds. B. Binggeli \& R. Buser, Springer
  Verlag, 317

\bibitem[1994]{mat94} Mather, J.C. et al.~1994, ApJ 420, 439
  
\bibitem[1969]{mat69} Matsuda, T., Sato, H., \& Takeda, H. 1969, Prog.
  Theor. Phys. 41, 219
  
\bibitem[1995]{par95} Partridge, R. 1995, in: 3K: The Cosmic Microwave
  Background Radiation, Cambridge University Press

\bibitem[1968]{pee68} Peebles, P.J.E. 1968, ApJ 153, 1

\bibitem[2002]{pee02} Peebles, P.J.E. 2002, astro-ph/0207347

\bibitem[1999]{per99} Perlmutter, S., et al.~1999, ApJ 517, 565
  
\bibitem[1994]{pfe94} Pfenniger, D., \& Combes, F. 1994, AA 285, 94
  
\bibitem[1993]{puy93} Puy, D., Alecian, G., Le Bourlot, J., L\'eorat,
  \& J., Pineau des For\^ets, G. 1993, AA 267, 337

\bibitem[1996]{puy96} Puy, D., \& Signore, M. 1996, AA 305, 371
  
\bibitem[1997]{puy97} Puy, D., \& Signore, M. 1997, New Ast 2, 299
  
\bibitem[2002]{raz02} Razoumov, A.O, Norman, M.L., Abel, T., \& Scott,
  D.  2002, ApJ 572, 695

\bibitem[1998]{rie98} Riess, A., et al.~1998, AJ 116, 1009
  
\bibitem[1993]{san93} Sanford S., \& Allamandola, L. 1993, ApJ 409,
  L65
  
\bibitem[1992]{sha92} Shapiro, P. 1992, in: Astrochemistry of Cosmic
  Phenomena, Singh P. ed., Kluwer Academic

\bibitem[1980]{sil80} Silvera, I.F. 1980, ``The solid molecular
  hydrogens in the condensed phase: fundamentals and static
  properties'', Rev. Mod. Phys. 52, 393

\bibitem[2000]{sri00} Srianand, R., Petitjean, P., \& Ledoux, C. 2000,
  Nature 408, 931

\bibitem[1996]{sta96} Stancil, P., Lepp, S., \& Dalgarno, A. 1996, ApJ
  458, 401

\bibitem[1998]{sta98} Stancil, P., Lepp, S., \& Dalgarno, A. 1998, ApJ
  509, 1
  
\bibitem[1969]{tak69} Takeda, H., Sato, H., \& Matsuda, T. 1969, Prog.
  Theor. Phys. 41, 840

\bibitem[2000]{tyt00} Tytler, D., O'Meara, J., \& Lubin, D. 2000,
  Physica Scripta 85, 12

\bibitem[1996]{uhe96} Uehara, H., Susa, H., Nishi, R., Yamada, M., \&
  Nakamura, T. 1996, ApJ 483, L95

\bibitem[2000]{uhe00} Uehara, H., \& Inutsuka, S. 2000, ApJ 531, L91
  
\bibitem[1935]{ure35} Urey, H. 1935, ``Some thermodynamic properties
  of hydrogen and deuterium'', in: Le Prix Nobel en 1934, Stockholm:
  Kungl. Baklryckenet, 1
 
\bibitem[1949]{van49} van de Hulst, H.C. 1949, Recherche Astron.
  Utrecht 11, 1
  
\bibitem[1987]{var87} Varghese, G., Prasad, R.D.G., \& Paddi Reddy, S.
  1987, Phys. Rev A. 35, 701

\bibitem[1999]{war99} Wardle, M., \& Walker, M. 1999, ApJ, 527, L109
  
\bibitem[1988]{wet88} Wetterich, C. 1988, Nucl. Phys. B 302, 321
  
\bibitem[1968]{wic68} Wickramasinghe, N., \& Reddish, V. 1968, Nature
  218, 661

\bibitem[1959]{zwi59} Zwicky, F. 1959, PASP 71, 468

\end{thebibliography}
\end{document}